\pdfoutput=1

\documentclass[11pt]{article}
\usepackage{acl}
\usepackage{multirow}
\usepackage{enumitem}
\usepackage{array}
\usepackage{latexsym}
\usepackage{graphicx}
\usepackage{subfigure}
\usepackage{enumitem}
\setlist[itemize]{itemsep=0pt, parsep=0pt, leftmargin = 0pt}
\usepackage{amsthm,amsmath,amssymb}
\usepackage{mathrsfs}
\usepackage{booktabs}
\usepackage{float}
\usepackage{booktabs}
\usepackage{multirow}
\usepackage{graphicx}
\usepackage{times}

\usepackage[T1]{fontenc}

\usepackage[utf8]{inputenc}

\usepackage{microtype}
\usepackage{inconsolata}
\title{Why Multi-Interest Fairness Matters: Hypergraph Contrastive Multi-Interest Learning for Fair Conversational Recommender System}

\author{{Yongsen Zheng}\textsuperscript{\textnormal{1,2}}\textbf{,} {Zongxuan Xie}\textsuperscript{\textnormal{3}}\textbf{,} {Guohua Wang}\textsuperscript{\textnormal{4}\textdagger}\textbf{,}\\
{\textbf{Ziyao Liu}}\textsuperscript{\textnormal{1,2}}\textbf{,} {\textbf{Liang Lin}}\textsuperscript{\textnormal{3,5}}\textbf{,} {\textbf{{Kwok-Yan Lam}}\textsuperscript{\textnormal{1,2}\textdagger}}\\
\textsuperscript{\textnormal{1}}Nanyang Technological University, Singapore\, \textsuperscript{\textnormal{2}}Digital Trust Centre Singapore\\
 \textsuperscript{\textnormal{3}}Sun Yat-sen University\, \textsuperscript{\textnormal{4}}South China Agricultural University\, \textsuperscript{\textnormal{5}}Peng Cheng Laboratory\\
 \{yongsen.zheng, liuziyao, kwokyan.lam\}@ntu.edu.sg\\
 xiezx25@mail2.sysu.edu.cn, wangguohua@scau.edu.cn, linliang@ieee.org \\
}

\begin{document}
\maketitle

\renewcommand{\thefootnote}{}
\footnotetext{\textsuperscript{\textnormal{\textdagger}}Corresponding author.
}

\begin{abstract}
{\color{black}
Unfairness is a well-known challenge in Recommender Systems (RSs), often resulting in biased outcomes that disadvantage users or items based on attributes such as gender, race, age, or popularity. Although some approaches have started to improve fairness recommendation in offline or static contexts, the issue of unfairness often exacerbates over time, leading to significant problems like the Matthew effect, filter bubbles, and echo chambers. To address these challenges, we proposed a novel framework, \textbf{Hy}pergraph Contrastive Multi-Interest Learning for \textbf{Fair} \textbf{C}onversational \textbf{R}ecommender \textbf{S}ystem (\textbf{HyFairCRS}), aiming to promote multi-interest diversity fairness in dynamic and interactive Conversational Recommender Systems (CRSs). HyFairCRS first captures a wide range of user interests by establishing diverse hypergraphs through contrastive learning. These interests are then utilized in conversations to generate informative responses and ensure fair item predictions within the dynamic user-system feedback loop. Experiments on two CRS-based datasets show that HyFairCRS achieves a new state-of-the-art performance while effectively alleviating unfairness. Our code is available at {\color{blue} https://github.com/zysensmile/HyFairCRS}.
}
\end{abstract}

\section{Introduction}
Conversational Recommender Systems (CRSs) have become an important research area \cite{HyCoRec, HiCore, qian2023hutcrs}, focusing on understanding user preferences through multi-turn dialogues. However, issues like data imbalance and inconsistent user-item interactions can result in unfair recommendations for disadvantaged users, leading to unsatisfactory suggestions. Thus, it is rather crucial to alleviate the unfairness in the dynamic and interactive CRS.

Recently, there has been a growing interest in examining fairness in Recommender Systems (RSs), which cater to multiple stakeholders \cite{two_sided_3}, including customers (user side) and providers (item side). These fairness-aware recommendation methods can be categorized into three main areas. The first category \cite{customer_side_1,customer_side_2,customer_side_3,customer_side_4} focuses on customer-side fairness, aiming to ensure equitable treatment of users across different groups (\emph{e.g.}, gender, race) to eliminate discrimination in recommendations. The second category \cite{provider_side_1,provider_side_2,provider_side_3,provider_side_4} addresses provider-side (item-side) fairness, ensuring relatively fair recommendation opportunities for diverse item groups (\emph{e.g.}, popular vs. unpopular) to create a fair platform for all providers. The third category \cite{two_sided_1,two_sided_2,two_sided_3,two_sided_4} investigates two-sided fairness, which seeks to mitigate bias in recommendations by ensuring that both users and items are treated equitably. This highlights the significance of balanced representation.

{\color{black}
Despite their effectiveness, most existing methods still encounter two major challenges. \emph{1) Unfairness Evolution.} Many methods aim to alleviate unfairness in offline or relatively static recommendation settings, often failing to account for the dynamic nature of unfairness in real-world scenarios. In practice, a feedback loop exists between users and the system, which can intensify unfairness over time as users engage with the recommendations. This escalation can result in significant issues, such as the Matthew effect \cite{MatthewEffect1}, filter bubbles \cite{filter_bubble}, and echo chambers \cite{echo_chamber}. Hence, it is crucial to address unfairness in the dynamic CRS. \emph{2) Diversity Fairness.} Some studies \cite{fairness_diversity_1,fairness_diversity_2} have indicated that when RSs negative fairness in user interest diversity and focus exclusively on relevance scores, they often produce recommendations that are overly similar or homogeneous. This lack of diversity can have negative consequences for both customers and providers. Customers may experience fatigue and boredom due to a surplus of similar items, which can diminish their long-term satisfaction \cite{fairness_diversity_1}. On the provider side, less popular items may struggle to gain visibility, overshadowed by a few dominant choices. Thus, investigating diversity fairness related to multiple interests is vital for fostering more balanced and equitable recommendations.}

\indent {\color{black}To address these issues, we propose a novel framework, \textbf{Hy}pergraph Contrastive Multi-Interest Learning for \textbf{Fair} \textbf{C}onversational \textbf{R}ecommender \textbf{S}ystem (\textbf{HyFairCRS}), aimed at exploring diversity fairness related to users' multiple interests in the interactive. HyFairCRS consists of two key components: Hypergraph Contrastive Multi-Interest Learning and Fair CRS. The former seeks to capture a wide range of user interests by creating diverse hypergraphs through contrastive learning. It starts by generating various hypergraphs, \emph{i.e.}, entity-, item-, word-, and review-guided, using different external knowledge sources (\emph{e.g.}, DBpedia \cite{DBpedia}, ConceptNet \cite{ConceptNet}, item reviews \cite{RevCore}) along with historical conversation data. Each hypergraph is decoupled into a hypergraph and a line graph, which are processed through hypergraph convolution and graph convolution layers to produce multiple representations of user interests. Finally, contrastive learning is used to refine these multi-interest embeddings, enhancing their diversity. The latter leverages these diverse interests to promote fairness in the CRS. Specifically, it integrates varied user interests into conversational tasks to generate informative responses and into recommendation tasks to boost result diversity, ensuring fairness across multi-turn dialogues. Extensive experiments on two widely-used CRS-based datasets show that HyFairCRS achieves a new state-of-the-art performance, validating its superiority of improving fairness.}

\indent Overall, our main contributions are included:
\vspace{-10pt}
\begin{itemize}[leftmargin=8pt]
\item To the best of our knowledge, this is the first work to investigate diversity fairness related to multiple interests in the dynamic CRS that involves a user-system feedback loop.
\item We proposed HyFairCRS, a novel end-to-end paradigm that aims to learn multiple user interests to generate informative responses in conversational tasks and to ensure fair item predictions in recommendation tasks.
\item Extensive experimental results on two CRS-based datasets show the effectiveness of HyFairCRS and its superiority of alleviating unfairness.
\end{itemize}

\section{Related Work}
\subsection{Conversational Recommender System}
{\color{black}Conversational Recommender System \cite{FacetCRS, CoMoRec, qian2023hutcrs, HyCoRec} assists users in finding products, services, or content through natural language interactions. It has two key components: the recommendation task, which matches items to user preferences, and the conversational task, which generates relevant responses. Currently, CRS-based methods fall into two groups: attributed-based CRS and generation-based CRS. The former \cite{deng2021unified, lei2020estimation,lei2020interactive,ren2021learning} focuses on leveraging specific attributes of users and items to provide personalized recommendations. These systems typically use structured data, such as user profiles (demographics, preferences) and item features (genre, price, ratings). The latter \cite{qian2023hutcrs,zhou2020improving,HiCore,shang2023multi} utilizes natural language generation techniques to create responses and recommendations dynamically. Instead of relying solely on predefined attributes, these systems can generate conversational responses based on context and dialogue history. 
While both approaches enhance user understanding, they struggle to alleviate unfairness in the CRS as user engagement evolves over time.}

\subsection{Fair Recommendation}
{\color{black}In the context of recommendations \cite{fairnessmetrics, TNNLS_1, TNNLS_2, music_recommendation, GCFM}, fair recommendation aims to ensure equitable outcomes for all users and items, minimizing biases related to sensitive attributes like gender, race, age, and socioeconomic status. The goal is to create systems that provide relevant recommendations while promoting fairness and preventing the reinforcement of existing inequalities. Fairness-aware recommendation systems can be categorized based on their stakeholder focus. 1) Customer-side fairness \cite{customer_side_1, customer_side_2, customer_side_3, customer_side_4}; 2) Provider-side fairness \cite{provider_side_1, provider_side_2, provider_side_4}, and 3) Two-sided fairness \cite{two_sided_1, two_sided_2, two_sided_4}. For example, \cite{customer_side_2} is proposed to enhance user embeddings with additional interaction data and implements a fairness strategy to improve recommendations for diverse user groups, and  \cite{provider_side_3} is devised to improve item exposure equity across groups via a fairness-constrained reinforcement learning algorithm, modeling it as a constrained Markov decision process. However, these methods overlook the negative impact of user-system feedback loop on fairness.}

\section{HyFairCRS}
\begin{figure*}[t]
    \centering
    \includegraphics[width=1\textwidth]{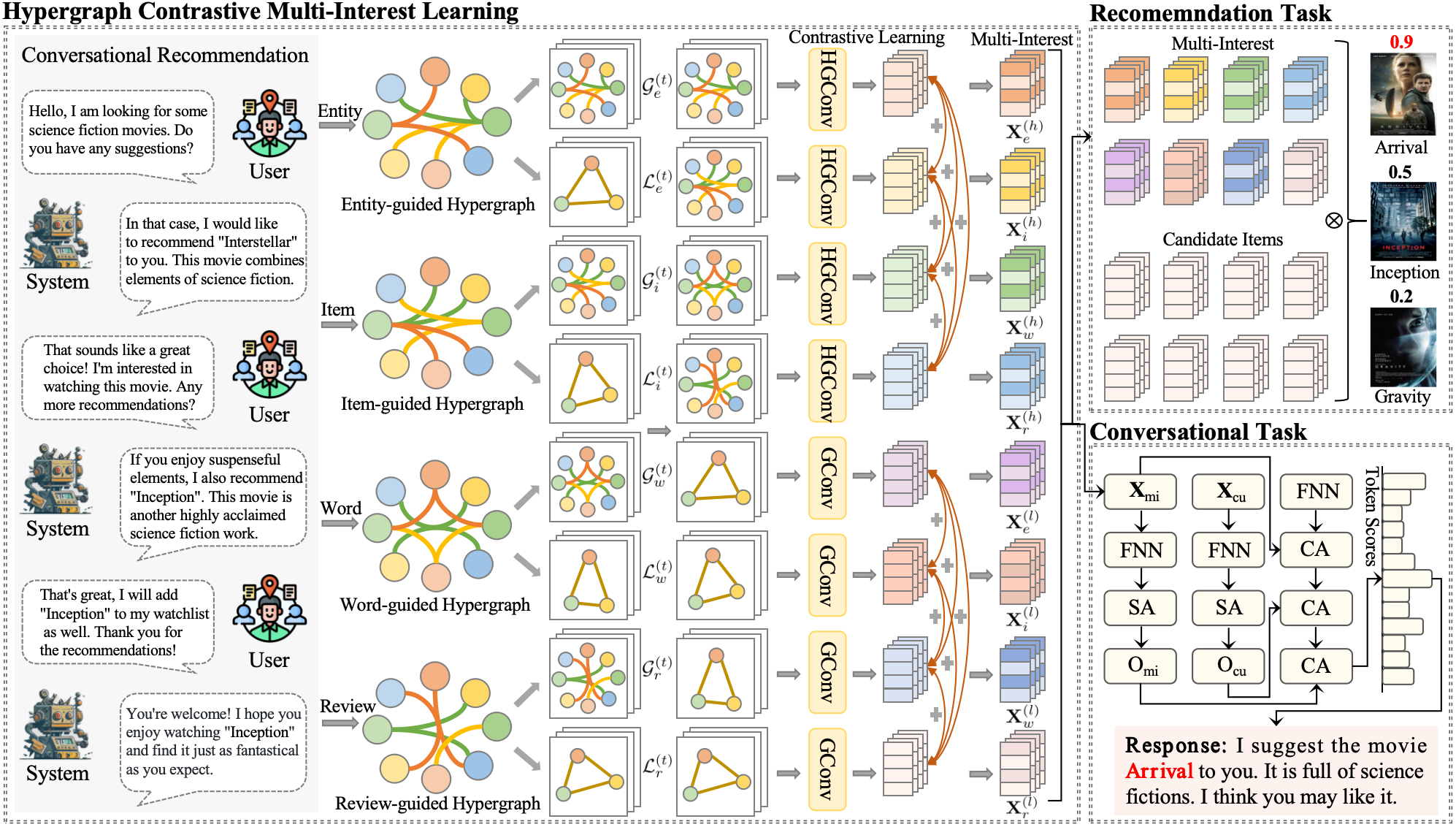} 
    \caption{Overview of our HyFairCRS, which consists of Hypergraph Contrastive Learning and Fair CRS. The former aims to learn multiple user interests to explore diversity fairness, while the latter adopts these diverse interests to generate responses in the conversation task, and to make fair item predictions in the recommendation task.}
    \label{fig:framework}
\end{figure*}

Enhancing fairness in the CRS is challenging, as unfairness can worsen over time through dynamic interactions. To this end, we propose HyFairCRS, which combines Hypergraph Contrastive Multi-Interest Learning with fair CRS. The pipeline of HyFairCRS is shown in Fig.\ref{fig:framework}.

\subsection{Preliminaries}

\noindent \textbf{Hypergraph.} A hypergraph is defined as \(\mathcal{G}^{(t)}_{\rm *} = (\mathcal{V}^{(t)}_{*}, \mathcal{E}^{(t)}_{*}, \textbf{H}^{(t)}_{*})\), where \(* \in \{ e, i, w, r \}\) represents entity, item, word, or review. Here, \(\mathcal{V}^{(t)}_{*}\), \(\mathcal{E}^{(t)}_{*}\), and \(\textbf{H}^{(t)}_{*}\) are the sets of nodes, hyperedges, and the weighted adjacency matrix, respectively. Let \( \boldsymbol{\rm V}^{(t)}_{*} \in \mathbb{N}^{|\mathcal{V}^{(t)}_{*}| \times |\mathcal{V}^{(t)}_{*}|} \) and \( \boldsymbol{\rm E}^{(t)}_{*} \in \mathbb{N}^{|\mathcal{E}^{(t)}_{*}| \times |\mathcal{E}^{(t)}_{*}|} \) be diagonal matrices of vertex and edge degrees.

\noindent \textbf{Line Graph.} A line graph for each hypergraph is written as \(\mathcal{L}^{(t)}_{\rm *} = (\widetilde{\mathcal{V}}^{(t)}_{*}, \widetilde{\mathcal{E}}^{(t)}_{*}, \widetilde{\textbf{H}}^{(t)}_{*})\) to capture hyperedge-level information. \(\widetilde{\mathcal{V}}^{(t)}_{*}\) corresponds to hyperedges from the original hypergraph, connected if they share a common node. \(\widetilde{\mathcal{E}}^{(t)}_{*}\) consists of pairs of hyperedges sharing at least one node, while \(\widetilde{\textbf{H}}^{(t)}_{*}\) is the incidence matrix representing relationships between hyperedges and their nodes.


\subsection{Hypergraph Contrative Multi-Interest Learning}
To model diverse preferences to address the unfairness issue \cite{fairness_diversity_1, customer_side_2}, we devise hypergraphs to effectively capture high-order user relationship patterns as a hypergraph improves traditional graphs by linking multiple vertices through hyperedges, allowing us to uncover and learn diverse user interests. 

\subsubsection{Multi-Hypergraph Construction}
To explore fairness related to multiple user interests, we establish a set of hypergraphs, including entity-, item-, word-, and review-guided hypergraphs.\\
\indent \emph{1) Entity-Guided Hypergraph:} We harness entities from historical sessions to extract multi-hop neighbors from the extensive DBpedia Knowledge Graph (KG) \cite{DBpedia}, enabling the construction of hyperedges linked by shared entities. Subsequently, the hyperedges are interconnected via common entities, culminating in the formation of an entity-guided hypergraph. \emph{2) Item-Guided Hypergraph:} 
It centers on items from previous conversations, treating them as nodes and forming hyperedges that connect through shared items. All hyperedges are linked together by the items in common to build item-guided hypergraph. This enables us to explore diverse users' interests towards different items. \emph{3) Word-Guided Hypergraph:} Keywords reflect users' dynamic needs, and we use items from conversations as keywords and identify their multi-hop neighbors in the ConceptNet KG \cite{ConceptNet}. This forms interconnected hyperedges, linked by common words to create the hypergraph. \emph{4) Review-Guided Hypergraph:} Item reviews highlight the importance of integrating user feedback and comments to improve recommendation performance. We utilize items from each conversation to search positive and negative words in the reviews, thereby creating distinct hyperedges. Formally, they can be given as:
\begin{equation}
\left\{
\begin{aligned}
\mathcal{G}^{(t)}_{e}&=(\mathcal{V}^{(t)}_{i} \cup \mathcal{V}^{(t)}_{n_{\rm d}}, \mathcal{E}^{(t)}_{e}, \textbf{H}^{(t)}_{e});\\
\mathcal{G}^{(t)}_{i}&=(\mathcal{V}^{(t)}_{i},\mathcal{E}^{(t)}_{i}, \textbf{H}^{(t)}_{i});\\
\mathcal{G}^{(t)}_{w}&=(\mathcal{V}^{(t)}_{i} \cup \mathcal{V}^{(t)}_{n_{c}},\mathcal{E}^{(t)}_{w}, \textbf{H}^{(t)}_{w});\\
\mathcal{G}^{(t)}_{r}&=(\mathcal{V}^{(t)}_{i} \cup \mathcal{V}^{(t)}_{n_{r}},\mathcal{E}^{(t)}_{r}, \textbf{H}^{(t)}_{r}).\\
\end{aligned}
\right.
\label{incidence_matrix}
\end{equation}
Here \(\mathcal{V}^{(t)}_{i}\), \(\mathcal{V}^{(t)}_{n_{\rm d}}\), \(\mathcal{V}^{(t)}_{n_{\rm c}}\), and \(\mathcal{V}^{(t)}_{n_{\rm r}}\) represent nodes from historical conversations, DBpedia KG, nodes from ConceptNet KG, and item reviews, respectively. Additionally, \(\mathcal{E}^{(t)}_{e}\), \(\mathcal{E}^{(t)}_{i}\), \(\mathcal{E}^{(t)}_{w}\), and \(\mathcal{E}^{(t)}_{r}\) denote the entity-, item-, word-, and review-guided hyperedge sets. The matrices \(\textbf{H}^{(t)}_{*}\) correspond to their incidence matrices, defined as follows:
\begin{equation}
\textbf{H}^{(t)}_{{*}_{(v, h)}} =  \left\{
\begin{aligned}
& 1, && {\rm if} \quad v \in h \\
& 0, && {\rm if} \quad v \notin h\\
\end{aligned}
\right.
\label{incidence_matrix}
\end{equation}For each node \( v \), its degree is given by \( d(v) = \sum_{h \in \mathcal{E}^{(t)}_{*}} \textbf{H}^{(t)}_{*_{(v, h)}} \); for each hyperedge \( h \), its degree is defined as \( \delta(h) = \sum_{v \in \mathcal{V}^{(t)}_{*}} \textbf{H}^{(t)}_{*_{(v, h)}} \). 

\subsubsection{Line Graph Induction} 
Since the hypergraph captures node-level information (intra-hyperedge), we create a corresponding line graph for each hypergraph to explore  hyperedge-level relationships (inter-hyperedge). This approach not only captures local messages but also explores global information. The interaction between these two types of information deepens our understanding of diverse user preferences. Formally, the line graphs associated with all hypergraphs can be expressed as:
\begin{equation}
\left\{
\begin{aligned}
\mathcal{L}^{(t)}_{e}&=(\widetilde{\mathcal{V}}^{(t)}_{i} \cup \widetilde{\mathcal{V}}^{(t)}_{n_{\rm d}}, \widetilde{\mathcal{E}}^{(t)}_{e}, \textbf{L}^{(t)}_{e});\\
\mathcal{L}^{(t)}_{i}&=(\widetilde{\mathcal{V}}^{(t)}_{i},\widetilde{\mathcal{E}}^{(t)}_{i}, \textbf{L}^{(t)}_{i});\\
\mathcal{L}^{(t)}_{w}&=(\widetilde{\mathcal{V}}^{(t)}_{i} \cup \widetilde{\mathcal{V}}^{(t)}_{n_{c}},\widetilde{\mathcal{E}}^{(t)}_{w}, \textbf{L}^{(t)}_{w});\\
\mathcal{L}^{(t)}_{r}&=(\widetilde{\mathcal{V}}^{(t)}_{i} \cup \widetilde{\mathcal{V}}^{(t)}_{n_{r}}, \widetilde{\mathcal{E}}^{(t)}_{r}, \textbf{L}^{(t)}_{r}).\\
\end{aligned}
\right.
\label{incidence_matrix}
\end{equation}Here \(\widehat{\mathcal{V}}^{(t)}_{*}\) contains nodes \(v_e\) for hyperedges \(e\) in \(\mathcal{E}^{(t)}_{*}\). The set \(\widehat{\mathcal{E}}^{(t)}_{*}\) includes pairs \((v_{e_p}, v_{e_q})\) for hyperedges \(e_p\) and \(e_q\) that share at least one node. The incidence matrix \(\textbf{L}^{(t)}_{*}\) defines the structure of \(\mathcal{L}^{(t)}_{\rm *}\), and we assign each edge \((v_{e_p}, v_{e_q})\) a weight $W_{p, q}$, which can be expressed as:
\begin{equation}
W_{p, q} = \frac{|e_p \cap e_q|}{|e_p \cup e_q|}.
\label{edge_wegith}
\end{equation}
\subsubsection{Contrastive Multi-Interest Learning}
Once we have generated a series of hypergraphs along with their corresponding line graphs, we shift our focus to examining the diverse interests of users through various contrastive learning paradigms. To streamline this exploration, we categorize user interests into two distinct types: hypergraph-oriented interests representations and line graph-oriented interest representations. To obtain these diverse representations, we input all hypergraphs (\emph{i.e.}, \(\mathcal{G}^{(t)}_{*}\)) into the Hypergraph Convolution Network (HGConv) to learn the intra-hyperedge representations. Simultaneously, we feed all line graphs (\emph{i.e.}, \(\mathcal{L}^{(t)}_{*}\)) into the Graph Convolution Network (GConv) to learn the inter-hyperedge representations. This dual strategy enables us to derive a set of primary feature representations for user interests, defined as:
\begin{equation}
\left\{
\begin{aligned}
&\textbf{O}^{(l+1)}_{*}  = \textbf{V}^{-1}_{*} \textbf{H}^{(t)}_{*} \textbf{W}^{(l)}_{*} \textbf{E}^{-1}_{*} {\textbf{H}^{(t)}_{*}}^T \textbf{O}^{(l)}_{*};\\
\textbf{X}^{(h)}_{*} &=\textbf{O}^{(L)}_{*}= \textbf{V}^{-1}_{*} \textbf{H}^{(t)}_{*} \textbf{W}^{(L-1)}_{*} \textbf{E}^{-1}_{*} {\textbf{H}^{(t)}_{*}}^T \textbf{O}^{(L-1)}_{*}.
\label{final_hypergraph_aware_preferences} 
\end{aligned}
\right.
\end{equation}
\begin{equation}
\left\{
\begin{aligned}
&\widehat{\textbf{O}}^{(l+1)}_{*}  = \sigma(\widehat{\textbf{V}}^{-\frac{1}{2}}_{*} \widehat{\textbf{N}}_{*} \widehat{\textbf{V}}^{-\frac{1}{2}}_{*} \widehat{\textbf{O}}^{(l)}_{*} {\Theta}^{(l)}_{\rm *});\\
\textbf{X}^{(l)}_{*}&=\widehat{\textbf{O}}^{(\widehat{L})}_{*}=\sigma(\widehat{\textbf{V}}^{-\frac{1}{2}}_{*} \widehat{\textbf{N}}_{*} \widehat{\textbf{V}}^{-\frac{1}{2}}_{*} \widehat{\textbf{O}}^{(\widehat{L}-1)}_{*} {\Theta}^{(\widehat{L}-1)}_{\rm *}).
\label{final_linegraph_aware_preferences} 
\end{aligned}
\right.
\end{equation}
In HGConv, \(\textbf{O}^{(l)}_{*}\) and \(\textbf{O}^{(l+1)}_{*}\) denote the output embeddings for layers \(l\) and \(l+1\), with \(\textbf{W}^{(l)}_{*}\) representing the trainable parameters. Information is aggregated from nodes to hyperedges using \({\textbf{H}^{(t)}_{*}}^T \textbf{O}^{(l)}_{*}\) and from hyperedges back to nodes through the premultiplication of \(\textbf{H}^{(t)}_{*}\). In GConv, the input and output feature matrices for layer \(l\) are \(\widehat{\textbf{O}}^{(l)}_{*}\) and \(\widehat{\textbf{O}}^{(l+1)}_{*}\). The adjacency matrix of the line graph is defined as \(\widehat{\textbf{N}}_{*} = \widehat{\textbf{H}}_{*} + \textbf{I}\), where \(\textbf{I}\) is the identity matrix. The degree matrix \(\widehat{\textbf{V}}^{-\frac{1}{2}}_{*}\) is computed with \(\widehat{\textbf{V}}_{{*}_{(p,p)}} = \sum^m_{q=1} \widehat{\textbf{N}}_{{*}_{(p,q)}}\), where \(m\) is the number of nodes per hyperedge. For simplicity, we consider the outputs from the final layers \(L\) and \(\widehat{L}\) as the hypergraph-oriented and line graph-oriented interest representations, respectively.

Next, we employ contrastive learning to refine both hypergraph-guided interest representations (\emph{i.e.}, \(\textbf{X}^{(h)}_{*}\)), and line graph-guided interest representations (\emph{i.e.}, \(\textbf{X}^{(l)}_{*}\)). The primary reason for adopting contrastive learning is its ability to enhance the robustness of learned representations, making them invariant to different views of the same instance by attracting positive pairs and separating negative pairs. Firstly, we employ the InfoNCE \cite{shang2023multi} contrastive loss to refine any pairs of hypergraph-guided interest representations as:
\begin{equation}
\mathcal{J}^{\mathcal{H}}_{(i,e)} = -\log \frac{ \exp( \text{sim}(\textbf{X}^{(h)}_{e}, {\textbf{X}^{(h)}_{i}}^{+} ) / \tau ) }{ \sum_{k=0}^{K} \exp ( \text{sim}(\textbf{X}^{(h)}_{e}, {\textbf{X}^{(k)}_{i}}^{-}) / \tau )},
\label{hypergraph_contrastive_learning} 
\end{equation} where \( \text{sim}(\cdot) \) and \( \tau \) are the cosine similarity function and temperature hyperparameter, respectively. \({\textbf{X}^{(h)}_{i}}^{+}\) is the positive sample matched with \({\textbf{X}^{(h)}_{i}}\), while \({\textbf{X}^{(k)}_{i}}^{-}\) refers to \(k\) negative samples. Analogously, we define the contrastive losses for other pairs of hypergraph-guided representations, namely \((\textbf{X}^{(h)}_{e}, \textbf{X}^{(h)}_{w})\), \((\textbf{X}^{(h)}_{e}, \textbf{X}^{(h)}_{r})\), \((\textbf{X}^{(h)}_{i}, \textbf{X}^{(h)}_{w})\), \((\textbf{X}^{(h)}_{i}, \textbf{X}^{(h)}_{r})\), and \((\textbf{X}^{(h)}_{w}, \textbf{X}^{(h)}_{r})\) as \(\mathcal{J}^{\mathcal{H}}_{(e,w)}\), \(\mathcal{J}^{\mathcal{H}}_{(e,r)}\), \(\mathcal{J}^{\mathcal{H}}_{(i,w)}\), \(\mathcal{J}^{\mathcal{H}}_{(i,r)}\), and \(\mathcal{J}^{\mathcal{H}}_{(w,r)}\), respectively. The total loss is obtained by summing the hypergraph contrastive losses across any two pairs:
\begin{equation}
\mathcal{J}^{\mathcal{H}}_{CL} = \mathcal{J}^{\mathcal{H}}_{(e,i)} + \mathcal{J}^{\mathcal{H}}_{(w,i)} + \mathcal{J}^{\mathcal{H}}_{(r,i)} + \mathcal{J}^{\mathcal{H}}_{(i,w)} + \mathcal{J}^{\mathcal{H}}_{(i,r)} + \mathcal{J}^{\mathcal{H}}_{(w,r)}.
\label{hypergraph_contrastive_total_loss_function} 
\end{equation}Similarly, we can derive the final contrastive loss for the line graph-based interest representations using Eq.(\ref{hypergraph_contrastive_learning}) and Eq.(\ref{hypergraph_contrastive_total_loss_function}) as \(\mathcal{J}^{\mathcal{L}}_{CL}\). To obtain the overall contrastive loss, we average the losses from both the hypergraph-based and line graph-based interest representations as:
\begin{equation}
\mathcal{J}_{CL} =  \mathcal{J}^{\mathcal{H}}_{CL}  + \mathcal{J}^{\mathcal{L}}_{CL}.
\label{hypergraph_linegraph_total_losses} 
\end{equation}
\subsection{Fair CRS}
To ensure fair recommendation, we next apply these interest representations into the CRS.

\subsubsection{Multi-Interest Fairness} 
Most existing methods \cite{fairness_diversity_1} indicate that exploring diversity fairness related to multiple user interests can mitigate unfairness. Thus, we integrate hypergraph- and line graph-guided interest representations into both recommendation and conversational tasks. We first concatenate all interest representations, then apply average pooling (${\rm Pooling}(\cdot)$) and Multi-Head Attention (${\rm MHA}(\cdot)$) \cite{Trans} to obtain the fair representations \(\textbf{X}_{\rm fair\_reco}\) and \(\textbf{X}_{\rm fair\_conv}\) for item prediction and response generation, respectively.
Formally, this procedure can be written as:
\begin{equation}
\begin{aligned}
&\textbf{X}_{\rm fair} = [\textbf{X}^{(h)}_{e};\textbf{X}^{(h)}_{i};\textbf{X}^{(h)}_{w}; \textbf{X}^{(h)}_{r};\textbf{X}^{(l)}_{e};\textbf{X}^{(l)}_{i};\textbf{X}^{(l)}_{w}; \textbf{X}^{(l)}_{r}],\\
&\textbf{X}_{\rm fair\_reco} = {\rm Pooling}([{\rm Pooling}(\textbf{X}_{\rm fair}); \textbf{X}_{\rm curr}]),\\
&\textbf{X}_{\rm fair\_conv} = {\rm MHA}(\textbf{X}_{\rm curr}, \textbf{X}_{\rm fair}, \textbf{X}_{\rm fair}),
\end{aligned}
\label{incidence_matrix}
\end{equation} where \( ; \) denote concatenation, and \(\textbf{X}_{\rm curr}\) represent current conversation representations. 
\subsubsection{Recommendation Task}
The recommendation task aims to suggest relevant items based on user preferences, enhancing satisfaction and engagement. To ensure fairness, we use \(\textbf{X}_{\rm fair\_reco}\) to make fair item prediction. Let \(\textbf{I}_{\rm cand}\) be the feature representations of candidate items in the set \(\mathcal{I}\). The recommendation probabilities for these items are given as:
\begin{equation}
\begin{aligned}
\textsf{P}_{\rm rec} = {\rm Softmax}(\textbf{X}_{\rm fair\_reco} \cdot \textbf{I}_{\rm cand}). 
\end{aligned}
\label{recommendation_probabilities}
\end{equation}
Then, we define the cross-entropy loss \(\mathcal{J}_{\rm R}\) as our main objective, supported by the contrastive loss \(\mathcal{J}_{\rm CL}\) to enhance recommendations. This allows us to effectively learn user preferences while ensuring diverse suggestions, as illustrated below:
\begin{equation}
\begin{aligned}
\mathcal{J}_{\rm R} = &- \sum^{\mathcal{B}}_{j=1} \sum^{|\mathcal{I}|}_{i=1} [y_{ij}{\rm log}({\textsf{P}}^{(j)}_{\rm rec}(i))\\
&+(1-y_{ij}){\rm log}(1-{\textsf{P}}^{(j)}_{\rm rec}(i))];
\end{aligned}
\end{equation}
\begin{equation}
\begin{aligned}
\mathcal{J}_{\rm CL\_R} = \alpha \mathcal{J}_{\rm CL} + \mathcal{J}_{\rm R},
\end{aligned}
\end{equation}
where \(\mathcal{B}\) is the mini-batch size, \(|\mathcal{I}|\) is the item set size, \(y_{ij} \in [0, 1]\) is the target label, and \(\alpha\) controls the contrastive loss impact.
\subsubsection{Conversational Task}
The conversational task is responsible for generating recommendation responses when the system chats with users via natural language dialogues. We incorporate fair conversation-based representations \(\textbf{X}_{\rm fair\_conv}\) in a Transformer-based encoder-decoder framework to generate informative and diverse responses. If \(\textbf{T}^{(n-1)}\) is the previous output, then the current output \(\textbf{T}^{(n)}\) is:
\begin{equation}
\begin{aligned}
\textbf{A}^{n}_0 &= {\rm MHA}(\textbf{T}^{n-1}, \textbf{T}^{n-1}, \textbf{T}^{n-1}),\\
\textbf{A}^{n}_1 &= {\rm MHA}(\textbf{A}^{n}_0, \textbf{X}_{\rm fair\_conv}, \textbf{X}_{\rm fair\_conv}),\\
\textbf{A}^{n}_2 &= {\rm MHA}(\textbf{A}^{n}_1, \textbf{X}_{\rm curr}, \textbf{X}_{\rm curr}),\\
\textbf{A}^{n}_3 &= {\rm MHA}(\textbf{A}^{n}_2, \textbf{X}_{\rm hist}, \textbf{X}_{\rm hist}),\\
\textbf{A}^{n}_4 &= \gamma \cdot \textbf{A}^{n}_2 + (1 - \gamma) \cdot \textbf{A}^{n}_3 ,\\
\textbf{T}^{n} &= {\rm FFN}(\textbf{A}^{n}_4),
\label{decoder}
\end{aligned}
\end{equation}where \(\textbf{X}_{\rm curr}\) and \(\textbf{X}_{\rm hist}\) represent the features of the current and historical conversations, \(\gamma\) regulates the information flow, and \({\rm FFN}(\cdot)\) denotes a fully-connected network. To enhance response diversity, the probability of generating the next token, given the previous sequence \(\{s_{t-1}\} = s_1, s_2, \ldots, s_{t-1}\), can be defined as:
\begin{equation}
\begin{aligned}
\textsf{P}_{\rm con}( s_t | \{s_{t-1}\}) = \sum^3_{j=1}\mathcal{P}_{j}(s_t | \mathcal{X}_j);
\label{conversation_loss_1}
\end{aligned}
\end{equation}
Here $\mathcal{X}_1$=$T_i$, $\mathcal{X}_2$=$\textbf{X}_{\rm fair\_conv}$, $\mathcal{X}_3 = (T_i, \textbf{X}_{\rm fair\_conv})$. \(T_i\) is the output generated by the decoder, while \(s_t\) denotes the \(t\)-th utterance in the sequence. \(\mathcal{P}_1(\cdot)\) is vocabulary probability based on the input \(T_i\), \(\mathcal{P}_2(\cdot)\) means vocabulary bias from \(\textbf{X}_{\rm fair\_conv}\), and \(\mathcal{P}_3(\cdot)\) indicates copy probability, with non-item vocabularies scored as zero. We train the model using cross-entropy loss with a contrastive loss as:
\begin{equation}
\begin{aligned}
\mathcal{J}_{\rm C} = - \sum_{b=1}^\mathcal{B} \sum_{t=1}^{\mathcal{T}} {\rm log}(\textsf{P}_{\rm con}( s_t | \{s_{t-1}\}));
\label{conversation_loss_2}
\end{aligned}
\end{equation}
\begin{equation}
\begin{aligned}
\mathcal{J}_{\rm CL\_C} =\beta \mathcal{J}_{\rm CL} + \mathcal{J}_{\rm C}.
\label{conversation_loss_2}
\end{aligned}
\end{equation}Here $\mathcal{T}$ is the truncated length of utterances, and $\beta$ is to control the contrastive loss.

\begin{table*}
\small
\setlength{\tabcolsep}{8mm}
\setlength{\abovecaptionskip}{4pt}  
\centering
\renewcommand{\arraystretch}{1.0}
\begin{tabular}{l@{\hskip 0.00in}
l@{\hskip 0.13in}
c@{\hskip 0.08in}c@{\hskip 0.08in}c@{\hskip 0.08in}c@{\hskip 0.08in}c@{\hskip 0.08in}c@{\hskip 0in}
c@{\hskip 0.18in}
c@{\hskip 0.08in}c@{\hskip 0.08in}c@{\hskip 0.08in}c@{\hskip 0.08in}c@{\hskip 0.08in}c@{\hskip 0in}
c@{\hskip 0.13in}}
\toprule
\multirow{2}{*}{\textbf{}}&
\multirow{2}{*}{\textbf{Model}}&
\multicolumn{6}{c}{REDIAL}&
&
\multicolumn{6}{c}{TG-REDIAL}
&\\
\cline{3-8}
\cline{10-15}
\rule{0pt}{10pt}
&&R@10&R@50&M@10&M@50&N@10&N@50&&R@10&R@50&M@10&M@50&N@10&N@50\\
&TextCNN&0.0644&0.1821&0.0235&0.0285&0.0328&0.0580&
&0.0097&0.0208&0.0040&0.0045&0.0053&0.0077\\
&SASRec&0.1117&0.2329&0.0540&0.0593&0.0674&0.0936&
&0.0043&0.0178&0.0011&0.0017&0.0019&0.0047\\
&BERT4Rec&0.1285&0.3032&0.0475&0.0555&0.0663&0.1045&
&0.0043&0.0226&0.0013&0.0020&0.0020&0.0058\\
&ReDial&0.1705&0.3077&0.0677&0.0738&0.0925&0.1222&
&0.0038&0.0165&0.0012&0.0017&0.0018&0.0045\\
&TG-ReDial&0.1679&0.3327&0.0694&0.0771&0.0924&0.1286&
&0.0110&0.0174&0.0048&0.0050&0.0062&0.0076\\
&KBRD&0.1796&0.3421&0.0722&0.0800&0.0972&0.1333&
&0.0201&0.0501&0.0077&0.0090&0.0106&0.0171\\
&KGSF&0.1785&0.3690&0.0705&0.0796&0.0956&0.1379&
&0.0215&0.0643&0.0069&0.0087&0.0103&0.0194\\
&BERT&0.1608&0.3525&0.0597&0.0688&0.0831&0.1255&
&0.0040&0.0194&0.0011&0.0017&0.0018&0.0050\\
&XLNet&0.1569&0.3590&0.0583&0.0677&0.0811&0.1255&
&0.0040&0.0187&0.0011&0.0017&0.0017&0.0048\\
&BART&0.1693&0.3783&0.0646&0.0744&0.0888&0.1350&
&0.0047&0.0187&0.0012&0.0017&0.0020&0.0048\\
&KGConvRec&0.1819&0.3587&0.0711&0.0794&0.0969&0.1358&
&0.0220&0.0524&0.0088&0.0102&0.0119&0.0185\\
&MHIM&0.1966&0.3832&0.0742&0.0830&0.1027&0.1440&
&0.0300&0.0783&0.0108&0.0129&0.0152&0.0256\\
&HiCore&0.2192&0.4163&0.0775&0.0874&0.1107&0.1558&
&0.0270&0.0769&0.0088&0.0107&0.0152&0.0225\\
&\textbf{HyFairCRS*}&\textbf{0.2237}&\textbf{0.4382}&\textbf{0.0798}&\textbf{0.0899}&\textbf{0.1297}&\textbf{0.1590}&
&\textbf{0.0358}&\textbf{0.0883}&\textbf{0.0156}&\textbf{0.0179}&\textbf{0.0174}&\textbf{0.0290}\\
\bottomrule
\end{tabular}
\caption{\label{tab:recommendation} Recommendation results. * indicates statistically significant improvement (\emph{p} < 0.05) over all baselines.}
\end{table*}

\subsection{Model Complexity Analysis} 
The complexity of HyFairCRS involves four main tasks. 1) Hypergraph construction: \(O(n)\); 2) Hypergraph convolution: \(O(n)\); 3) Contrastive learning: \(O(n \times d)\); 4) Line graph construction: \(O(n^2 \times e)\) where \(n\), \(d\) are batch size and feature dimension.

The first three tasks have \textbf{linear complexity}, and we optimized the fourth task to achieve \(O(n k \log k)\) through: a) Set Operation Optimization: Using NumPy's `intersect1d' and `union1d' reduced `Jaccard' index computation from \(O(e)\) to \(O(k \log k)\); b) JIT Compilation: Numba's `njit' enhances speed; c) Loop Parallelization: `prange' enables multi-core execution. After optimization, the theoretical complexity of HyFairCRS is \(O\left(\frac{n^2 k \log k}{p}\right)\), where \(p\) is the number of CPU cores. On multi-core machines, \(p\) can often approach \(n\) (\emph{e.g.}, on our 128-core machine), simplifying the complexity to \(O\left(nk \log k\right)\). Compared with the strongest baseline \textbf{HiCore} \cite{HiCore}, has a complexity of \(O(n^3)\) because it requires identifying all distinct triangle motifs to construct hyperedges, involving the enumeration of all triplets. Our method outperforms HiCore in both accuracy and efficiency. 

\section{Experiments and Analyses}
To fully evaluate our HyFairCRS, we conduct experiments to answer the following questions:
\vspace{-8pt}
\begin{itemize}
\item \textbf{RQ1:} How does HyFairCRS perform compared with all baselines in the recommendation task?
\item \textbf{RQ2:} How does HyFairCRS perform compared with all baselines in the conversation task?
\item \textbf{RQ3:} How does HyFairCRS alleviate the unfairness in the CRS?
\item \textbf{RQ4:} How does each type of hypergraph contribute to the performance?
\item \textbf{RQ5:} How does HyFairCRS perform on the cross-domain datasets?
\item \textbf{RQ6:} How do parameters affect our HyFairCRS?
\end{itemize}

\subsection{Experimental Protocol}
\textbf{Datasets.}
We evaluate our HyFairCRS on two challenging public CRS-based benchmarks: \textbf{REDIAL} \cite{TDCR} and \textbf{TG-REDIAL} \cite{Topic-Guided}. Reviews of REDIAL and TG-REDIAL are sourced from IMDb\footnote{https://www.dbpedia.org/} and Douban\footnote{https://movie.douban.com/}, respectively. To further evaluate our proposed HyFairCRS, we incorporate additional cross-domain datasets, including \textbf{OpenDialKG} \cite{opendialkg} and \textbf{DuRecDial} \cite{durecdial}. These datasets encompass a variety of domains, such as movies, music, sports, books, news, culinary experiences, and restaurants.\\
\noindent \textbf{Baselines.} To fully evaluate our HyFairCRS, we compared with a series of state-of-the-art methods, including \textbf{TextCNN} \cite{TextCNN}, \textbf{SASRec} \cite{SASRec}, \textbf{BERT4Rec} \cite{BERT4Rec}, \textbf{Transformer} \cite{Trans}, \textbf{ReDial} \cite{li2018towards}, \textbf{KBRD} \cite{chen2019towards}, \textbf{KGSF} \cite{zhou2020improving}, \textbf{KGConvRec} \cite{sarkar2020suggest}, \textbf{BERT} \cite{BERT}, \textbf{XLNet} \cite{XLNet}, \textbf{BART} \cite{BART}, \textbf{DialoGPT} \cite{DialoGPT}, \textbf{GPT-3} \cite{GPT3}, \textbf{C2-CRS} \cite{C2CRS}, \textbf{LOT-CRS} \cite{LOTCRS}, \textbf{HiCore} \cite{HiCore}, \textbf{UniCRS} \cite{UniCRS}, \textbf{MHIM} \cite{shang2023multi}, and \textbf{HyCoRec} \cite{HyCoRec}.

\subsection{Recommendation Performance (RQ1)}
For the recommendation task, we use several metrics: Recall@\emph{K} (R@\emph{K}), Mean Reciprocal Rank@\emph{K} (MRR@\emph{K}), and Normalized Discounted Cumulative Gain@\emph{K} (NDCG@\emph{K}) for \emph{K}=10 and 50 \cite{shang2023multi,HiCore}. HyFairCRS consistently outperforms all baselines, as shown in Table \ref{tab:recommendation}. Its superior performance is attributed to: 1) Diverse Hypergraphs: our HyFairCRS constructs hypergraphs that include various dimensions (emph{i.e.}, entities, words, items, and reviews) for capturing the complexity of user interests and enabling the identification of intricate relationships for more relevant suggestions. 2) Decoupled Structure: each hypergraph is split into a hypergraph and a line graph. The hypergraph focuses on intra-hyperedge relationships, while the line graph examines inter-hyperedge connections, enhancing flexibility and effectively capturing complex user-item interactions. 3) Multiple Interest Representations : the model features eight distinct interest representations, each targeting specific user preferences, leading to more personalized recommendations and a better user experience.

\subsection{Conversational Performance (RQ2)}
For the conversation task, we also employ several metrics \cite{CoMoRec}, including Distinct \(n\)-gram (Dist-\(n\)), BLEU-\(n\) for \(n \in [2, 3, 4]\). Table \ref{tab:conversation} demonstrate that HyFairCRS consistently outperforms all baselines. Key observations include:1) LLM-guided and KG-based methods (\emph{e.g.}, DialoGPT, GPT-3, MHIM, CoMoRec, HyCoRec, HiCore) generally perform better due to their ability to leverage rich knowledge for learning interest representations; 2) Token-sequence-aware approaches (\emph{e.g.}, Transformer, ReDial) show the weakest results, as they focus solely on token sequences and miss user preferences embedded in entities. 3) HyFairCRS outperforms debiasing-based methods (\emph{e.g.}, HyCoRec) by utilizing multiple external large-scale KGs and item reviews to create multidimensional hypergraphs (entity-, item-, word-, and review-guided). This approach effectively captures diverse user interests, enhancing recommendation diversity and enriching dialogue content and responses.

\begin{table}[t]
\scriptsize
\setlength{\tabcolsep}{0.1mm}
\setlength{\abovecaptionskip}{2pt}  
\centering
\renewcommand{\arraystretch}{1.1}
\begin{tabular}{l@{\hskip 0.04in}
c@{\hskip 0.04in}c@{\hskip 0.04in}c@{\hskip 0.03in}c@{\hskip 0in}
c@{\hskip 0.03in}
c@{\hskip 0.04in}c@{\hskip 0.04in}c@{\hskip 0.03in}c@{\hskip 0in}
c@{\hskip 0.04in}}
\hline
\multirow{2}{*}{\textbf{Model}}&
\multicolumn{4}{c}{REDIAL}&
&
\multicolumn{4}{c}{TG-REDIAL}
&\\
\cline{2-5}
\cline{7-10}
\rule{0pt}{10pt}
&Dist-2&Dist-3&BLEU-2&BLEU-3&
&Dist-2&Dist-3&BLEU-2&BLEU-3&\\
\hline
ReDial&0.0214&0.0659&0.0198&0.0054&
&0.2178&0.5136&0.0387&0.0094&\\
Trans.&0.0538&0.1574&0.0164&0.0027&
&0.2362&0.7063&0.0335&0.0075&\\
KBRD&0.0765&0.3344&0.0203&0.0061&
&0.8013&1.7840&0.0411&0.0107&\\
KGSF&0.0572&0.2483&0.0211&0.0067&
&0.3891&0.8868&0.0442&0.0128&\\
GPT-3&0.3604&0.6399&0.0225&0.0090&
&1.2255&2.5713&0.0438&0.0124&\\
DialoGPT&0.3542&0.6209&0.0220&0.0085&
&1.1881&2.4269&0.0440&0.0126&\\
UniCRS&0.2464&0.4273&0.0218&0.0080&
&0.6252&2.2352&0.0438&0.0126&\\
C2-CRS&0.2623&0.3891&0.0223&0.0088&
&0.5235&1.9961&0.0434&0.0120&\\
LOT-CRS&0.3312&0.6155&0.0222&0.0085&
&0.9287&2.4880&0.0428&0.0123&\\
MHIM&0.3278&0.6204&0.0226&0.0089&
&1.1100&2.3520&0.0435&0.0118&\\
CoMoRec&0.3492&0.6480&0.0230&0.0089&
&1.1273&2.5063&0.0437&0.0125&\\
HyCoRec&0.3661&0.6434&0.0228&0.0092&
&1.2590&2.6000&0.0440&0.0126&\\
HiCore&0.5871&1.1170&0.0231&0.0093&
&2.8610&5.7440&0.0441&0.0125&\\
\textbf{HyFairCRS*}&\textbf{0.6022}&\textbf{1.2270}&\textbf{0.0233}&\textbf{0.0098}&
&\textbf{2.9390}&\textbf{5.8360}&\textbf{0.0445}&\textbf{0.0128}&\\
\hline
\end{tabular}
\caption{\label{tab:conversation} Conversation results. * indicates statistically significant improvement (\emph{p} < 0.05) over all baselines.}
\end{table}

\begin{table*}
\scriptsize
\setlength{\tabcolsep}{2mm}
\setlength{\abovecaptionskip}{2pt}  
\centering
\renewcommand{\arraystretch}{1.1}
\begin{tabular}{l@{\hskip 0.1in}
c@{\hskip 0.05in}c@{\hskip 0.05in}c@{\hskip 0.05in}c@{\hskip 0.05in}c@{\hskip 0.05in}c@{\hskip 0.05in}c@{\hskip 0.05in}c@{\hskip 0in}
c@{\hskip 0.15in}
c@{\hskip 0.05in}c@{\hskip 0.05in}c@{\hskip 0.05in}c@{\hskip 0.05in}c@{\hskip 0.05in}c@{\hskip 0.05in}c@{\hskip 0.05in}c@{\hskip 0in}
c@{\hskip 0.1in}}
\hline
\multirow{2}{*}{\textbf{Model}}&
\multicolumn{8}{c}{REDIAL}&
&
\multicolumn{8}{c}{TG-REDIAL}
&\\
\cline{2-9}
\cline{11-18}
\rule{0pt}{10pt}
&A@5$\downarrow$&A@10$\downarrow$&A@15$\downarrow$&A@20$\downarrow$&G@5$\downarrow$&G@10$\downarrow$&G@15$\downarrow$&G@20$\downarrow$&
&A@5$\downarrow$&A@10$\downarrow$&A@15$\downarrow$&A@20$\downarrow$&G@5$\downarrow$&G@10$\downarrow$&G@15$\downarrow$&G@20$\downarrow$&\\
\hline
KBRD&0.0062&0.0099&0.0131&0.0158&0.8282&0.8494&0.8533&0.8534&
&0.0181	&0.0236&	0.0291&	0.0325	&0.8265&	0.8156	&0.8020	&0.7948&\\  
KGSF&0.0062&0.0103&0.0135&0.0166&0.8291&0.8472&0.8529&0.8518&
&0.0188&	0.0246	&0.0291&	0.0321&	0.8504&	0.8494&	0.8413&	0.8370&\\  
KGConvRec&0.0063&0.0099&0.0132&0.0162&0.8366&0.8534&0.8297&0.8539&
&0.0161&	0.0227	&0.0265&	0.0303&	0.8681&	0.8328	&0.8188&	0.8063&\\   
MHIM&0.0038&0.0056&0.0071&0.0084&0.7962&0.8198&0.8569&0.8323&
&0.0024&	0.0034	&0.0043	&0.0051&	0.6858&	0.7078&	0.7129&	0.7185&\\    
HyCoRec&0.0030&0.0045&0.0057&0.0068&0.7653&0.7961&0.8063&0.8117&
&0.0012	&0.0017&	0.0021&	0.0025&	0.5325&	0.5789&	0.5993&	0.6105&\\   
HiCore&0.0031&0.0045&0.0056&0.0067&0.7836&0.8129&0.8254&0.8302&
&0.0011	&0.0017	&0.0021&	0.0026&	0.4892&	0.5380	&0.5617&	0.5734&\\   
\textbf{HyFairCRS}&\textbf{0.0028}&\textbf{0.0040}&\textbf{0.0051}&\textbf{0.0061}&\textbf{0.7163}&\textbf{0.7504}&\textbf{0.7645}&\textbf{0.7711}&
&\textbf{0.0010}&\textbf{0.0015}&\textbf{0.0018}&\textbf{0.0022}&\textbf{0.4825}&\textbf{0.5313}&\textbf{0.5545}&\textbf{0.5666}&\\
\hline						
&L@5$\downarrow$&L@10$\downarrow$&L@15$\downarrow$&L@20$\downarrow$&D@5$\uparrow$&D@10$\uparrow$&D@15$\uparrow$&D@20$\uparrow$&
&L@5$\downarrow$&L@10$\downarrow$&L@15$\downarrow$&L@20$\downarrow$&D@5$\uparrow$&D@10$\uparrow$&D@15$\uparrow$&D@20$\uparrow$&\\
\hline
KBRD&5.1706&5.6387	&5.8862	&6.0446&	0.3967&	0.3890	&0.3908&	0.5557&
&5.5726&	5.7429	&5.8589	&5.9245&	0.2262	&0.1759	&0.2102	&0.2594&\\   
KGSF&5.1796	&5.6642	&5.9170&	6.0768	&0.3540	&0.3509&	0.3819	&0.4010&
&5.7729&	6.0155&	6.1143&	6.1734&	0.1794&	0.1796&	0.2092&	0.2558&\\   
KGConvRec&5.2627	&5.6834	&5.9335&6.0751&	0.3535&	0.3649&	0.3692	&0.5320&
&5.8173	&5.8702&	5.9034&	5.9428&	0.1259&	0.1501&	0.1932	&0.2366&\\  
MHIM&4.5794	&5.0229&	5.2573&	5.4124&	0.6067&	0.6499	&0.6700&	0.6624&
&3.0119	&3.3942&	3.6156	&3.7780&	0.5490&	0.6106&	0.6385&	0.6632&\\   
HyCoRec&4.1947	&4.6710&	4.9217&	5.0967&	0.6198&	0.6634&	0.6754	&0.6802&
&1.8647&	2.2999&	2.5614&	2.7640&	0.6814&	0.7099&	0.7394	&0.7548&\\   
HiCore&4.3901	&4.8461&	5.0915&	5.2607&	0.6472&	0.6804&	0.6870&	0.7002&
&1.6797&	2.1458&	2.4333	&2.6492&	0.6871&	0.7137&	0.7359	&0.7484&\\   
\textbf{HyFairCRS}&\textbf{3.5946}&\textbf{4.0402}&\textbf{4.2999}&\textbf{4.4765}&\textbf{0.6359}&\textbf{0.6813}&\textbf{0.7142}&\textbf{0.7249}&
&\textbf{1.5995}&\textbf{2.0094}&\textbf{2.2756	}&\textbf{2.4718	}&\textbf{0.7055}&\textbf{0.7341}&\textbf{0.7606}&\textbf{0.7751}&\\			
\hline
\end{tabular}
\caption{\label{tab:fairness_results} Fair recommendation comparisons via four fairness-aware metrics. Lower A@K, G@K, L@K values indicate greater fairness, while a higher D@K value signifies a fairer recommendation.}
\end{table*}

\begin{table}[h]
\scriptsize
\setlength{\tabcolsep}{0.1mm}
\setlength{\abovecaptionskip}{2pt}  
\centering
\renewcommand{\arraystretch}{1.1}
\begin{tabular}{l@{\hskip 0.04in}
c@{\hskip 0.03in}c@{\hskip 0.03in}c@{\hskip 0.03in}c@{\hskip 0in}
c@{\hskip 0.1in}
c@{\hskip 0.03in}c@{\hskip 0.03in}c@{\hskip 0.03in}c@{\hskip 0in}
c@{\hskip 0.03in}}
\hline
\multirow{2}{*}{\textbf{Model}}&
\multicolumn{4}{c}{REDIAL}&
&
\multicolumn{4}{c}{REDIAL}
&\\
\cline{2-5}
\cline{7-10}
\rule{0pt}{10pt}
&A@5$\downarrow$&A@10$\downarrow$&G@5$\downarrow$&G@10$\downarrow$&
&L@5$\downarrow$&L@10$\downarrow$&D@5$\uparrow$&D@10&$\uparrow$\\
\hline
\textbf{Ours*}&\textbf{0.0028}&\textbf{0.0040}&\textbf{0.7163}&\textbf{0.7504}&
&\textbf{3.5946}&\textbf{4.0402}&\textbf{0.6359}&\textbf{0.6813}&\\
w/o $\mathcal{G}^{(t)}_{e}$ &0.0034 &0.0050&0.7441 &0.7808&
&3.9215&4.375&0.5540	&0.6348&\\
w/o $\mathcal{G}^{(t)}_{i}$&0.0034 &0.0049 &0.7457 &0.7802&
&3.9138&4.3647&0.5573&0.6327\\
w/o $\mathcal{G}^{(t)}_{w}$&0.0034&0.0049	&0.7455&0.7806&
&3.9133&4.3691&0.5602&0.6305&\\
w/o $\mathcal{G}^{(t)}_{r}$&0.0036&0.0052&0.7490&0.7839&
&4.0019&4.4320&0.5671&0.6185&\\
w/o $\mathcal{L}^{(t)}_{e}$&0.0034&0.0049&0.7447&0.7808&
&3.9216&4.3701&0.5673&0.6426&\\
w/o $\mathcal{L}^{(t)}_{i}$&0.0034&0.0050&0.7460&0.7807&
&3.9138&4.3702&0.5608&0.6267&\\
w/o $\mathcal{L}^{(t)}_{w}$&0.0034&0.0050	&0.7447&0.7802&
&3.9124&4.3686&0.5617&0.6291&\\
w/o $\mathcal{L}^{(t)}_{r}$&0.0035&0.0050&0.7469&0.7830&
&3.9673&4.3905&0.5721&0.6395&\\
\hline
\multirow{2}{*}{\textbf{Model}}&
\multicolumn{4}{c}{TG-REDIAL}&
&
\multicolumn{4}{c}{TG-REDIAL}
&\\
\cline{2-5}
\cline{7-10}
\rule{0pt}{10pt}
&A@5$\downarrow$&A@10$\downarrow$&G@5$\downarrow$&G@10$\downarrow$&
&L@5$\downarrow$&L@10$\downarrow$&D@5$\uparrow$&D@10&$\uparrow$\\
\hline
\textbf{Ours*}&\textbf{0.0010}&\textbf{0.0015}&\textbf{0.4825}&\textbf{0.5313}&
&\textbf{1.5995}&\textbf{2.0094}&\textbf{0.7055}&\textbf{0.7341}&\\
w/o $\mathcal{G}^{(t)}_{e}$ &0.0013	&0.0018	&0.5253	&0.5671&
&1.8841	&2.2918&0.6726&0.7132&\\
w/o $\mathcal{G}^{(t)}_{i}$&0.0013&0.0019&0.5375&0.5787&
&1.9514	&2.3702&0.6570&0.6917&\\
w/o $\mathcal{G}^{(t)}_{w}$&0.0013	&0.0018	&0.5266	&0.5702&
&1.8790&	2.3144&0.6725&0.7028&\\
w/o $\mathcal{G}^{(t)}_{r}$&0.0013&0.0018&0.5190&0.5622&
&1.8549&2.2863&0.6756&0.7090&\\
w/o $\mathcal{L}^{(t)}_{e}$&0.0012&0.0018&	0.5225&	0.5646&	
&1.8598	&2.2795&	0.6689	&0.7089&\\
w/o $\mathcal{L}^{(t)}_{i}$&0.0013	&0.0018	&0.5285	&0.5706&
&1.9003	&2.3253&	0.6609&	0.6973&\\
w/o $\mathcal{L}^{(t)}_{w}$&0.0013	&0.0018	&0.5213	&0.5655&
&1.8613	&2.2885&0.6678&0.6981&\\
w/o $\mathcal{L}^{(t)}_{r}$&0.0013&0.0018&0.5164&0.5572&
&1.8452&2.2682&0.6643&0.7054&\\
\hline
\end{tabular}
\caption{\label{tab:ablation_studies} Ablation studies on the recommendation task.}
\end{table}


\subsection{Study on Fairness (RQ3)}
To assess if HyFairCRS reduces unfairness, we use several fairness metrics \cite{fairnessmetrics}: Average Popularity (A@\emph{K}), Gini Coefficient (G@\emph{K}), KL-Divergence (L@\emph{K}), and Difference (D@\emph{K}). Lower values for A@\emph{K}, G@\emph{K}, and L@\emph{K} indicate better fairness, while a higher D@\emph{K} suggests fairer recommendations. Table \ref{tab:fairness_results} shows our method achieves lower A@\emph{K}, G@\emph{K}, and L@\emph{K} values, with a higher D@\emph{K} compared to other baselines, indicating improved fairness.
This success can be attributed to several key factors: 1) HyFairCRS constructs eight distinct user interest representations, which offer deeper insights into users' preferences and effectively mitigate preference bias. This multifaceted approach allows for a more nuanced understanding of user behavior. 2) By decoupling the hypergraph into hypergraphs and line graphs, we are able to learn both node-level and hyperedge-level features. This dual perspective enhances preference diversification, allowing the model to capture a broader range of user interests and interactions. 3) We employ multi-set graph contrastive learning to refine user interest features. This technique helps to distinguish between similar user preferences, improving the model's ability to make accurate and relevant recommendations. These factors contribute to the superior performance of HyFairCRS, enabling it to adapt effectively across various domains and user contexts.

\begin{figure}[t]
    \centering
    \includegraphics[width=0.45\textwidth]{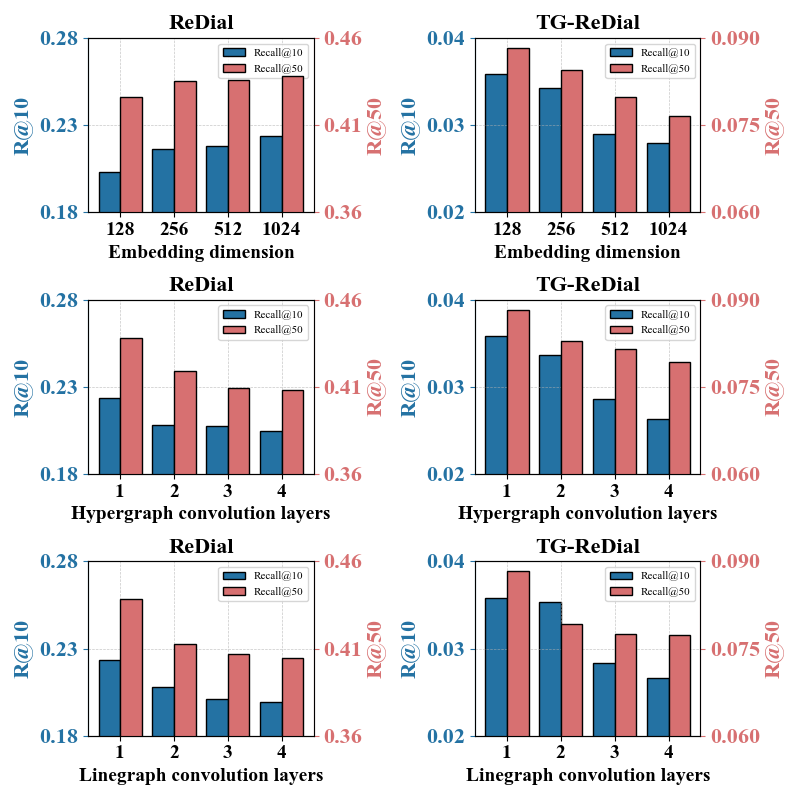} 
    \caption{Hyperparameter Analysis.}
    \label{fig:hyperparameters}
\end{figure}

\subsection{Ablation Studies (RQ4)}
To verify the contributions of each component in our HyFairCRS, we conduct ablation experiments, including: 1) remove different hypergraphs: w/o $\mathcal{G}^{(t)}_{e}$, w/o $\mathcal{G}^{(t)}_{i}$, w/o $\mathcal{G}^{(t)}_{w}$, w/o $\mathcal{G}^{(t)}_{r}$; 2) remove different line graphs: w/o $\mathcal{L}^{(t)}_{e}$, w/o $\mathcal{L}^{(t)}_{i}$, w/o $\mathcal{L}^{(t)}_{w}$, w/o $\mathcal{L}^{(t)}_{r}$. 
The results of these ablation studies, summarized in Table \ref{tab:ablation_studies}, reveal a consistent trend: removing any type of hypergraph leads to a significant decrease in performance metrics. This outcome confirms the critical role that each hypergraph plays in enhancing the model's overall effectiveness. For instance, the removal of the entity hypergraph may hinder the model's ability to accurately represent and relate different entities within the dataset, which is essential for generating contextually relevant recommendations. Similarly, excluding the user hypergraph could diminish the model's understanding of user preferences and behaviors, thereby impacting the personalization aspect of recommendations. Overall, these ablation experiments not only validate the individual contributions of each component in HyFairCRS but also highlight the synergistic effects that arise from their combined use, ultimately leading to a more robust and effective recommendation system. 

\subsection{Study on Cross-domain Datasets (RQ5)}
To thoroughly evaluate our proposed HyFairCRS, we incorporate two additional datasets that span diverse domains, including movies, music, books, sports, news, and restaurants. Table \ref{tab:Recommendation_opendialkg_durecdial} summarizes the experimental results. Notably, our method consistently outperforms the strongest baselines in both conversational and recommendation tasks across these cross-domain datasets. This consistent performance across varied domains underscores the robustness and scalability of HyFairCRS. The model's ability to adapt effectively to different contexts and user preferences is a significant advantage, as it allows for more personalized and relevant recommendations. For instance, whether users are seeking movie suggestions or looking for the latest news articles, HyFairCRS demonstrates a remarkable capacity to tailor its outputs to meet individual needs. Furthermore, the statistically significant improvements highlighted by an asterisk (*) reinforce the effectiveness of our approach in enhancing both recommendation accuracy and conversational quality. Overall, the results from these diverse datasets confirm that HyFairCRS is not only effective in its primary tasks but also possesses the versatility to handle a wide range of user interactions across different domains.

\begin{table}
\small
\setlength{\tabcolsep}{4mm}
\setlength{\abovecaptionskip}{6pt}  
\centering
\renewcommand{\arraystretch}{1.1}
\begin{tabular}{l@{\hskip 0.1in}
c@{\hskip 0.05in}c@{\hskip 0.05in}
c@{\hskip 0.15in}
c@{\hskip 0.05in}c@{\hskip 0.05in}
c@{\hskip 0.1in}}
\hline
\multirow{2}{*}{\textbf{Model}}&
\multicolumn{2}{c}{OpenDialKG}&
&
\multicolumn{2}{c}{DuRecDial}
&\\
\cline{2-3}
\cline{5-6}
\rule{0pt}{10pt}
&R@1&R@10&
&R@1&R@10&\\
\hline
KBRD&0.1448&0.3162&
&0.0618&0.3971&\\
KGSF&0.0626&0.1757&
&0.1395&0.4367&\\
ReDial&0.0008&0.0134&
&0.0005&0.0336&\\
TGReDial&0.2149&0.4035&
&0.0956&0.4882&\\
HyCoRec&0.2742&0.4490&
&0.1279&0.4750&\\
HiCore&0.2628&0.4526&
&0.1735&0.5471\\
\textbf{HyFairCRS*}&\textbf{0.2895}&\textbf{0.4630}&
&\textbf{0.1952}&\textbf{0.5688}\\
\hline
\multirow{2}{*}{\textbf{Model}}&
\multicolumn{2}{c}{OpenDialKG}&
&
\multicolumn{2}{c}{DuRecDial}
&\\
\cline{2-3}
\cline{5-6}
\rule{0pt}{10pt}
&Dist-2&Dist-3&
&Dist-2&Dist-3&\\
\hline
KBRD&0.3192&1.7660&
&0.5180&1.5500&\\
KGSF&0.1687&0.5387&
&0.1389&0.3862&\\
ReDial&0.1579&0.5808&
&0.1095&0.3981&\\
TGReDial&0.4836&2.1430&
&0.5453&2.0030&\\
HyCoRec&2.8190&4.7710&
&1.0820&2.4440&\\
HiCore&2.8430&4.8120&
&1.0940&2.4280\\
\textbf{HyFairCRS*}&\textbf{2.9162}&\textbf{5.0194}&
&\textbf{1.1210}&\textbf{2.4688}\\
\hline
\end{tabular}
\caption{\label{tab:Recommendation_opendialkg_durecdial} {\color{black}Results on both recommendation and conversation tasks on OpenDialKG and DuRecDial datasets involving various domains. * indicates statistically significant improvement (\emph{p} < 0.05) over all baselines.}}
\end{table}

\subsection{Hyperparameters Analysis (RQ6)}

We analyze the impact of key hyperparameters on overall performance (see Fig. \ref{fig:hyperparameters}). We can see that: 1) As we increase the number of hypergraph and line graph convolution layers in both datasets, we observe a decline in overall performance. This suggests that a single convolution layer is sufficient for effectively capturing the underlying patterns in the data. Additional layers may lead to overfitting, where the model becomes too tailored to the training data and struggles to generalize to unseen examples. 2) In the ReDial dataset, higher feature dimensions correlate with improved performance, indicating that more complex representations can enhance the model's ability to capture intricate user behaviors and preferences. In contrast, in the TG-ReDial dataset, we see a decline in performance with increased feature dimensions. This suggests that smaller datasets may benefit from simpler representations, as more complex features can introduce noise and reduce the model's effectiveness. More complex datasets, like ReDial, require higher dimensions to adequately represent the diversity of user interests and interactions. 

\section{Conclusion}
To enhance recommendation fairness in the CRS, we propose HyFairCRS, a novel paradigm that learns diverse interest representations through hypergraphs and line graphs during user interactions via natural language dialogues. These representations are used for generating responses in the conversation task and ensuring fair recommendations in the recommendation task. Extensive experiments show that HyFairCRS outperforms strong baselines in reducing unfair recommendations.

\section{Limitations}
Despite the promising results of HyFairCRS, two limitations warrant consideration. First, while HyFairCRS aims to promote fairness, it may not fully eliminate all biases inherent in the data. The model's effectiveness is contingent on the quality and representativeness of the training data; if the data reflects existing biases, these may persist in the recommendations. Second, user satisfaction and engagement are influenced by factors beyond fairness, such as relevance and personalization. Balancing these aspects while maintaining fairness remains a critical challenge for future research.

\section{Ethics Statement}
The data used in this paper are sourced from open-access repositories, and do not pose any privacy concerns. We are confident that our research adheres to the ethical standards set forth by ACL.

\section{Acknowledgements}
This research is supported by the National Research Foundation, Singapore and Infocomm Media Development Authority under its Trust Tech Funding Initiative. Any opinions, findings and conclusions or recommendations expressed in this material are those of the author(s) and do not reflect the views of National Research Foundation, Singapore and Infocomm Media Development Authority.



\bibliography{custom}  

\end{document}